\documentclass[journal=nalefd, manuscript=article]{achemso}

\usepackage{graphicx}
\usepackage{dcolumn}
\usepackage{bm}
\usepackage{braket} 

\title{Controlling Thermal Conductivity of Two-dimensional Materials via Externally Induced Phonon-Electron Interaction}

\author{Sheng-Ying Yue}
\affiliation{Department of Mechanical Engineering, University of California, Santa Barbara, CA 93106, USA}

\author{Runqing Yang}
\affiliation{Department of Mechanical Engineering, University of California, Santa Barbara, CA 93106, USA}

\author{Bolin Liao}
\email{bliao@ucsb.edu}
\affiliation{Department of Mechanical Engineering, University of California, Santa Barbara, CA 93106, USA}


\begin{document}

\begin{abstract}

Phonon scattering by electrons, or ``phonon-electron scattering", has been recognized as a significant scattering channel for phonons in materials with high electron concentration, such as thermoelectrics and nanoelectronics, even at room temperature. Despite the abundant previous studies of phonon-electron scattering in different types of three-dimensional (3D) bulk materials, its impact on the phonon transport, and thus the heat transfer properties, of two-dimensional (2D) materials has not been understood. In this work, we apply \textit{ab initio} methods to calculate the phonon-electron scattering rates in two representative 2D materials, silicene and phosphorene, and examine the potential of controlling the thermal conductivity of these materials via externally induced phonon-electron scattering by electrostatic gating. We also develop an analytical model to explain the impact of reduced dimensionality and distinct electron and phonon dispersions in 2D on phonon-electron scattering processes. We find that over 40\% reduction of the lattice thermal conductivity can be achieved in silicene with an induced charge carrier concentration in the range of $10^{13} \ \rm{cm}^{-2}$, which is experimentally achievable. Our study not only generates new fundamental insights into phonon transport in 2D materials but also provides practical guidelines to search for 2D materials with strong phonon-electron scattering for potential thermal switching applications. 

\end{abstract}

\maketitle


Electron-phonon interactions play a major role in determining the electronic properties of materials since they are the major contributors to electrical resistance and also mediate conventional superconductivity\cite{ziman2001electrons,grimvall1981electron}. Due to these reasons, the influence of electron-phonon interaction on the transport of electrons has been intensively studied and well understood. On the other hand, the scattering of phonons due to electron-phonon interactions (hereafter ``phonon-electron scattering") and its impact on thermal transport of solids have received limited research interest, due to the long belief that it is only important at cryogenic temperatures\cite{makinson1938thermal,klemens1954lattice,butler1978electron}. The main reason is that most of the previous studies and practical interests were limited to devices with a low or moderate electron concentration, typically below $10^{19}\ \mathrm{cm}^{-3}$. Recent technological developments have led to important applications with electron concentration as high as $10^{20}$ to $10^{21}\ \mathrm{cm}^{-3}$, such as in heavily-doped thermoelectric materials\cite{zhang2013high} and nanoelectronic devices\cite{del2011nanometre}. In this regime, however, the impact of phonon-electron scattering on thermal transport in largely unknown. Recently, Liao et al. used \textit{ab initio} calculations\cite{giustino2017electron} to show that the lattice thermal conductivity of silicon with a high electron concentration can be suppressed by as much as 50\% even at room temperature due to phonon-electron scattering\cite{liao2015significant}. Significant suppression of phonon propagation by phonon-electron scattering in silicon at room temperature was subsequently verified experimentally using ultrafast photoacoustic spectroscopy\cite{liao2016photo}. Moreover, \textit{ab initio} calculations of phonon-electron scattering have also been carried out in bulk metals\cite{jain2016thermal,wang2016first}, providing new insights into the details of coupled transport of phonons and electrons. 

\begin{figure*}[ht]
\includegraphics[width=1.0\textwidth,clip]{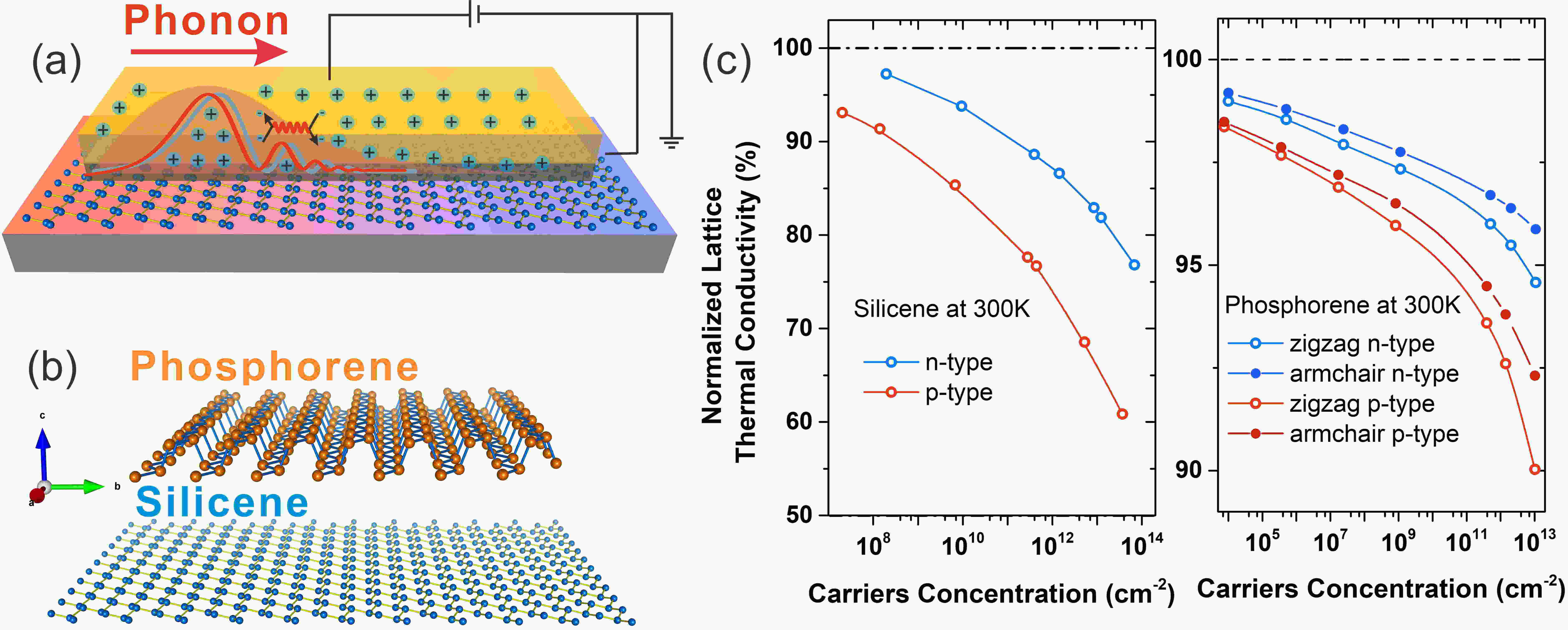}
\centering
\caption{(a) The concept of controlling phonon transport in 2D materials via externally induced phonon-electron scattering. (b) The crystal structures of phosphorene and silicene. (c) The calculated reduction of the lattice thermal conductivity of silicene and phosphorene (along zigzag and armchair directions) at 300\ K as a function of induced charge carrier concentration.}
\label{fig:1}
\end{figure*}

With the rapid advancement of nanotechnology, 2D materials have become star candidates for a wide range of applications, e.g. transistors, optoelectronics and energy harvesting devices. Currently, theoretical and experimental understanding of electron-phonon interaction in 2D materials has been limited to its influence on electrons and charge carrier mobility\cite{park2008electron,liao2015ab,efetov2010controlling}. The effect of phonon-electron scattering on the phonon frequency renormalization of 2D materials has been reported\cite{pisana2007breakdown}, but its effect on phonon lifetime and transport properties is still lacking. Given the paramount importance of phonon transport and thermal management for device performance in these applications, it is desirable to gain an in-depth understanding of phonon-electron scattering and its impact on phonon transport in 2D materials. From a fundamental point of view, phonon-electron scattering in 2D materials is expected to be qualitatively different from that in 3D bulk materials, due to factors including reduced dimensionality and thus altered scattering phase space, dominant normal phonon scattering and hydrodynamic phonon transport\cite{lee2015hydrodynamic,yang2019hydrodynamic}, new symmetries and the associated scattering selection rules\cite{lindsay2010flexural}, distinct electron and phonon dispersion relations (e.g. massless Dirac fermions and quadratic flexural phonons) and different dielectric screening behavior for polar materials\cite{yue2018ultralow}. Therefore, it is of fundamental interest to explore the behavior of phonon-electron scattering in 2D materials. From a practical point of view, thanks to the possibility of inducing a high concentration of electrons or holes in 2D materials via electrostatic gating\cite{efetov2010controlling,wu2018gate}, efficient phonon-electron scattering in 2D materials can potentially enable the development of fast thermal switches whose thermal conductivity can be tuned by applying an external electric field. This concept is illustrated in Fig.\ \ref{fig:1}a. Reversible and wide-range control of the thermal conductivity of solids using external fields is highly desirable in diverse fields\cite{wehmeyer2017thermal}. Previously, a variety of approaches have been explored experimentally to reversibly control the thermal conductivity of solids. For example, the metal-insulator phase transition\cite{kim2016thermal} and the associated crystal lattice change, e.g. of vanadium dioxide, has been proposed as a potential mechanism of thermal switching, but only a small contrast was achieved experimentally\cite{oh2010thermal}. In another example, a thermal conductivity tuning of 11\% was demonstrated by modifying the ferroelectric domain structure using an external electric field\cite{ihlefeld2015room}. Reversible electrochemical intercalation of ions has been demonstrated to change the thermal conductivity of layered materials\cite{zhu2016tuning}, but requires operation in a liquid electrolyte. Here we provide another possible scheme. With the development of solid-state electrolytes as the gate dielectric, inducing a charge concentration as high as $10^{14}\ \mathrm{cm}^{-2}$ in 2D materials\cite{efetov2010controlling,wu2018gate} has recently been demonstrated. In principle, this high density of charge carriers can efficiently scatter phonons and largely reduce the lattice thermal conductivity while an optimum charge carrier density can be selected to balance the increase of the electronic thermal conductivity. 

In this work, we use \textit{ab initio} electron-phonon interaction calculations\cite{giustino2017electron} to examine the practicality of this mechanism as a means to realize efficient thermal switching of 2D materials. We choose silicene and phosphorene as model systems due to their wrinkled crystal structures that break the out-of-plane mirror symmetry, as shown in Fig.\ \ref{fig:1}b. It has been theoretically shown that the out-of-plane mirror symmetry, e.g. of graphene, prohibits the first-order interaction of electrons with the out-of-plane flexural phonons\cite{castro2010limits,park2014electron}, which are the major heat carriers in 2D materials\cite{lindsay2010flexural}. Therefore, 2D materials without the out-of-plane mirror symmetry should be better candidates to possess strong phonon-electron scattering. Another reason to choose silicene and phosphorene is their distinct electronic structures: silicene is a Dirac semimetal with massless electrons and holes near the intrinsic Fermi level\cite{vogt2012silicene} and phosphorene is a semiconductor with a sizable band gap and highly anisotropic carrier effective masses\cite{fei2014enhanced,liao2015ab}. Therefore, a comparative study will help elucidate the impact of electronic structure on phonon-electron scattering and provide guidelines in future search for 2D materials with desirable phonon-electron scattering properties. We find that the lattice thermal conductivity of silicene and phosphorene depends strongly on the induced charge carrier concentration, as illustrated in Fig.\ \ref{fig:1}c. In particular, in p-type silicene, over 40\% reduction of the lattice thermal conductivity can be achieved by a charge carrier density of $10^{13} \mathrm{cm}^{-3}$, whereas in phosphorene, 10\% change of the lattice thermal conductivity can be expected at a similar charge carrier density. We provide in-depth analysis of these results in the following sections.  

The scattering rate of a phonon mode with wave vector $\mathbf{q}$ and branch label $\nu$ due to phonon-electron scattering is given by the Fermi's golden rule in the form\cite{grimvall1981electron,liao2015significant}:
\begin{equation} \label{eqn:1}
    \frac{1}{\tau^{\mathrm{ep}}_{\mathbf{q}\nu}}=-\frac{2\pi}{\hbar} \sum_{mn,\mathbf{k}} |g^{\nu}_{mn}(\mathbf{k},\mathbf{q})|^2 (f_{n\mathbf{k}}-f_{m(\mathbf{k}+\mathbf{q}+\mathbf{G})}) \delta(\epsilon_{n\mathbf{k}}-\epsilon_{m(\mathbf{k}+\mathbf{q}+\mathbf{G})}-\omega_{\mathbf{q}\nu})
\end{equation}
where $\tau^{\mathrm{ep}}_{\mathbf{q}\nu}$ is the corresponding phonon relaxation time, $g^{\nu}_{mn}(\mathbf{k},\mathbf{q})$ is the scattering matrix element connecting the initial electronic state with band index $n$ and wave vector $\mathbf{k}$ with the final electronic state with band index $m$ and wave vector $\mathbf{k}+\mathbf{q}+\mathbf{G}$ ($\mathbf{G}$ is a reciprocal lattice vector) due to the conservation of crystal momentum, $f_{n\mathbf{k}}$ is the Fermi-Dirac distribution, $\epsilon_{n\mathbf{k}}$ is the electron energy, and $\omega_{\mathbf{q}\nu}$ is the phonon frequency. The $\delta$ function imposes energy conservation during the scattering process. From Eq. \ref{eqn:1}, there are two major factors that contribute to the phonon-electron scattering rate: the scattering matrix elements that reflect the strength of the coupling between the electronic and lattice degrees of freedom and the energy-momentum conservation conditions that determine the number of potential scattering channels. To carry out the calculation in Eq. \ref{eqn:1}, the electronic band structure, the phonon dispersion and electron-phonon scattering matrix element are required. We use the density functional theory (DFT) based method to calculate these ingredients. The electronic band structure is calculated using the Quantum ESPRESSO package\cite{giannozzi2017advanced}. A mesh grid of $30 \times 30 \times 1$ in the first Brillouin zone is adopted for both materials with norm-conserving pseudo-potentials. The kinetic energy cutoff for wavefunctions is set to 40~Ry. The kinetic energy cutoff for charge density and potential is set to 160~Ry. The total electron energy convergence threshold for self-consistency is $\rm 1 \times 10^{-10}~Ry$. Applying density functional perturbation theory (DFPT) implemented in the same package, the phonon dispersion and the electron-phonon matrix elements are calculated on a coarse mesh of $6 \times 6 \times 1$ and $4 \times 4 \times 1$ for silicence and phosphorene, respectively. The results are subsequently interpolated onto a fine mesh of $60 \times 60 \times 1$ for silicene and $64 \times 64 \times 1$ for phosphorene using a maximally localized Wannier function based scheme as implemented in the EPW package\cite{jesse2010epw}. The calculated electronic band structure and phonon dispersion agree well with the literature and are given in the Supporting Information.

\begin{figure*}[ht]
\includegraphics[width=1.0\textwidth,clip]{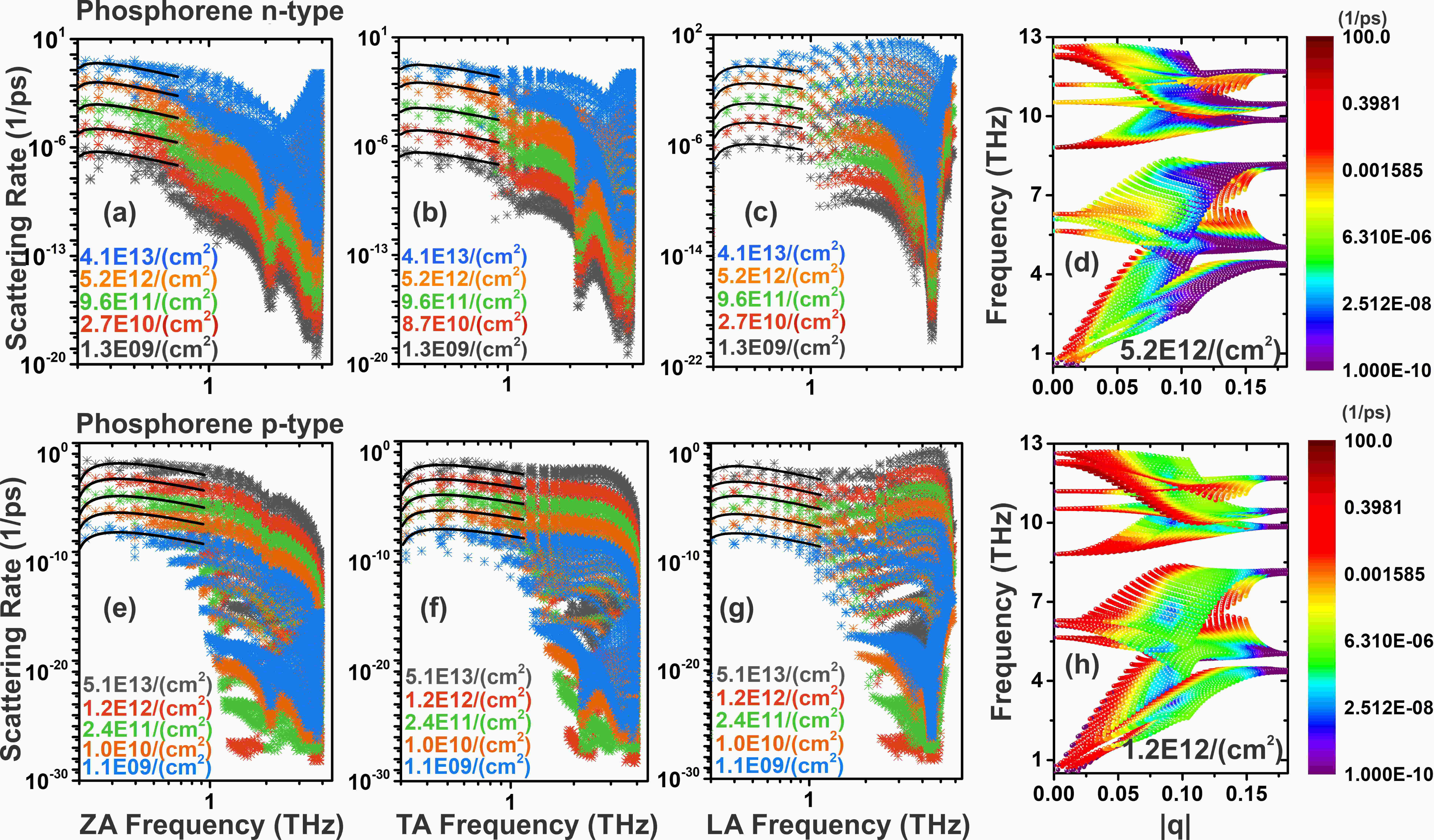}
\caption{Phonon-electron scattering rates of (a) ZA, (b) TA, and (c) LA modes as a function of phonon frequency at different carrier concentrations in n-type phosphorene. The solid lines are fits based on Eqns.~\ref{eqn:2} and \ref{eqn:3}. (d) Mode-resolved phonon-electron scattering rates in n-type phosphorene at a carrier concentration of $5.2 \times 10^{12} \rm{cm}^{-2}$. (e-f) are corresponding plots for p-type phosphorene.}
\label{fig:2}
\end{figure*}

Figure \ref{fig:2} shows the scattering rates of different phonon modes due to phonon-electron scattering in phosphorene. Both n-type and p-type phosphorene with different charge carrier concentrations are investigated. Several observations can be made compared to the results in 3D materials, such as bulk silicon\cite{liao2015significant}. While the phonon scattering rates scale linearly with the carrier concentration in both silicon and phosphorene, a major difference is the phonon frequency dependence of the scattering rates for the low-frequency acoustic phonons. In 3D semiconductors like silicon, the linear phonon dispersion and parabolic electronic bands give rise to the linear dependence of the phonon scattering rates on the phonon frequency for the low-frequency acoustic phonons\cite{liao2015significant}, whereas in phosphorene, the frequency dependence is more complicated: the scattering rates first rise with phonon frequency below 0.3~THz and then decrease. The decreasing trend of the phonon scattering rates in phosphorene indicates that the major heat-carrying phonons, typically around 1~THz, are less effectively scattered by electrons than phonons with longer wavelengths. To qualitatively understand the distinct phonon frequency dependence of the phonon-electron scattering rates in phosphorene, we develop an analytical model for the phonon-electron scattering rates of low-frequency acoustic phonons with the deformation potential approximation, where the electron-phonon matrix elements    $|g^{\nu}_{mn}(\mathbf{k},\mathbf{q})|^2$ are replaced by $\frac{\hbar D^2 q^2}{2 m_0 \omega_{\bf{q} \nu}}$ with $D$ being the constant deformation potential\cite{lundstrom2009fundamentals} and $m_0$ the atomic mass. The deformation potential is the change of the electron energy near the band edges in response to a static strain of the lattice, corresponding to the coupling strength of electrons with long-wavelength acoustic phonons. Using the phonon dispersion relations in 2D, including the quadratic dispersion for out-of-plane flexural phonons, and the 2D electronic band structure, we derive the following analytical equations for the phonon-scattering rates of ZA phonons in 2D semiconductors in the nondegenrate regime:
\begin{equation} \label{eqn:2}
    \frac{1}{\tau^{\mathrm{ep}}_{\mathbf{q}\nu}}=\frac{D^2(2\pi m^{*})^{\frac{1}{2}}}{\rho \alpha^{\frac{1}{2}} (k_\mathrm{B}T)^{\frac{3}{2}}} e^{-\frac{\hbar^2 (\frac{1}{2}+\frac{\alpha m^{*}}{\hbar})^2 \omega_{\mathbf{q}\nu}}{2m^{*}\alpha k_{\mathrm{B}} T}} n(E_\mathrm{F}) \omega^{\frac{1}{2}}_{\mathbf{q}\nu},
\end{equation}
where $m^{*}$ is the average electron or hole effective mass, $\rho$ is the areal mass density, $\alpha$ is the coefficient for the quadratic ZA phonon dispersion ($\omega=\alpha q^2$), $k_\mathrm{B}$ is the Boltzmann constant and $n(E_\mathrm{F})$
is the carrier concentration corresponding to a certain Fermi level $E_{\mathrm{F}}$. The detailed derivation is given in the Supporting Information. Similarly, we also derive an analytical expression for the phonon scattering rates of TA and LA phonon modes with linear dispersion relations given by:
\begin{equation} \label{eqn:3}
    \frac{1}{\tau^{\mathrm{ep}}_{\mathbf{q}\nu}}=\frac{D^2(2\pi m^{*})^{\frac{1}{2}}}{\rho v_{\mathrm{s}} (k_\mathrm{B}T)^{\frac{3}{2}}} e^{-\frac{\hbar^2 \omega_{\mathbf{q}\nu}^2}{8m^{*} v_{\mathrm{s}}^2 k_{\mathrm{B}} T}-\frac{\hbar\omega_{\mathbf{q}\nu}}{2k_{\mathrm{B}}T}-\frac{m^{*} v_\mathrm{s}^2}{2k_\mathrm{B}T}} n(E_\mathrm{F}) \omega_{\mathbf{q}\nu},
\end{equation}
where $v_\mathrm{s}$ is the speed of sound for the TA and LA phonon modes. These analytical results are plotted in Fig.~\ref{fig:2} as solid lines and agree very well with the \textit{ab initio} calculations, and confirm that the different phonon frequency dependency originates from the difference in dimensionality and the electron and phonon dispersion relations. These formulas can be used to estimate the level of phonon-electron scattering rates for long-wavelength acoustic phonons in 2D semiconductors. In Fig.~\ref{fig:2}(d) and (h), we also report the phonon-mode-resolved phonon-electron scattering rates in n-type and p-type phosphorene, where each data point in the plots represents a phonon mode with given frequency and wavevector magnitude and the color of the data point encodes the phonon-electron scattering rate of this mode. In both n-type and p-type phosphorene, phonons near the zone center are much more strongly scattered by electrons. This behavior is also observed in bulk silicon\cite{liao2015significant} and is expected due to the small energy scale of phonons compared to that of electrons, so that the small-wavevector phonons have a larger chance of meeting the energy-momentum conservation conditions during the scattering process. Furthermore, some of the zone boundary phonons, particularly in p-type phosphorene, are also significantly scattered by electrons. This is related to the strong anisotropy and large carrier effective mass along the zigzag direction in phosphorene\cite{fei2014enhanced,liao2015ab}, which leads to large radius of the electron and hole pockets near the band edges such that the electrons can be scattered across the carrier pockets by the zone boundary phonons with a large momentum change but a small energy change.   
\begin{figure}[ht]
\includegraphics[width=1.0\textwidth,clip]{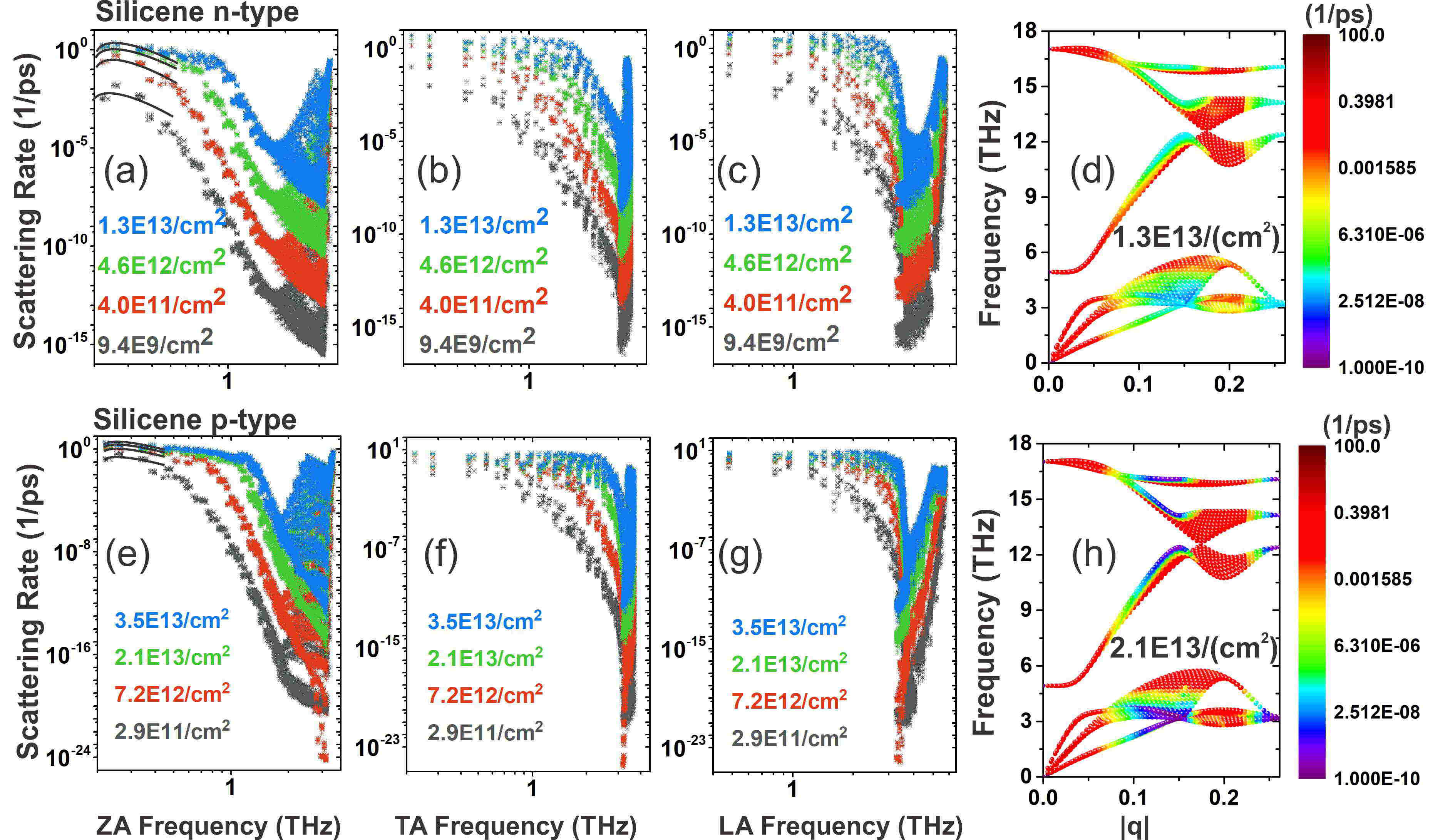}
\caption{Phonon-electron scattering rates of (a) ZA, (b) TA, and (c) LA modes as a function of phonon frequency at different carrier concentrations in n-type silicene. The solid lines are fits based on numerical integration of Eqn.~\ref{eqn:1} with the deformation potential approximation. (d) Mode-resolved phonon-electron scattering rates in n-type silicene at a carrier concentration of $1.3 \times 10^{13} \rm{cm}^{-2}$. (e-f) are corresponding plots for p-type silicene.}
\label{fig:3}
\end{figure}

In Fig.~\ref{fig:3}, we show the calculated scattering rates of phonons due to phonon-electron scattering in silicene. The main difference between silicene and phosphorene is their electronic structures: phosphorene is a wide-gap semiconductor and silicene is a Dirac semimetal with linear electron bands. The phonon frequency dependence of the scattering rates of ZA mode shows similar behavior as that in phosphorene, but the scattering rates are higher in silicene at a given carrier concentration in general. As the carrier concentration increases, the phonon scattering rates also exhibit a trend of saturation as the Fermi level moves deeper into the bands. Due to the linear electron bands and the degenerate nature requiring the use of Fermi-Dirac distribution functions in Eq.~\ref{eqn:1}, there is no simple analytical expression for the phonon scattering rates as those given for phosphorene (Eq.~\ref{eqn:2} and Eq.~\ref{eqn:3}) even with the deformation potential approximation. Instead, we integrate Eq.~\ref{eqn:1} numerically with the deformation potential approximation, and the results are shown as solid lines in Fig.~\ref{fig:3}, where good agreement with the \textit{ab initio} results is observed and the saturation of the scattering rates with increasing carrier concentration is captured. Figure \ref{fig:3}(d) and (h) display the mode-resolved phonon-electron scattering rates in n-type and p-type silicene. Similarly as in phosphorene, phonons near the zone center are strongly scattered by electrons. Phonons in a section of the Brillouin zone near the boundary are also strongly scattered, which is caused by intervalley scattering of electrons and holes between the two Dirac cones of silicene.

\begin{figure}[h!]
\includegraphics[width=0.6\textwidth,clip]{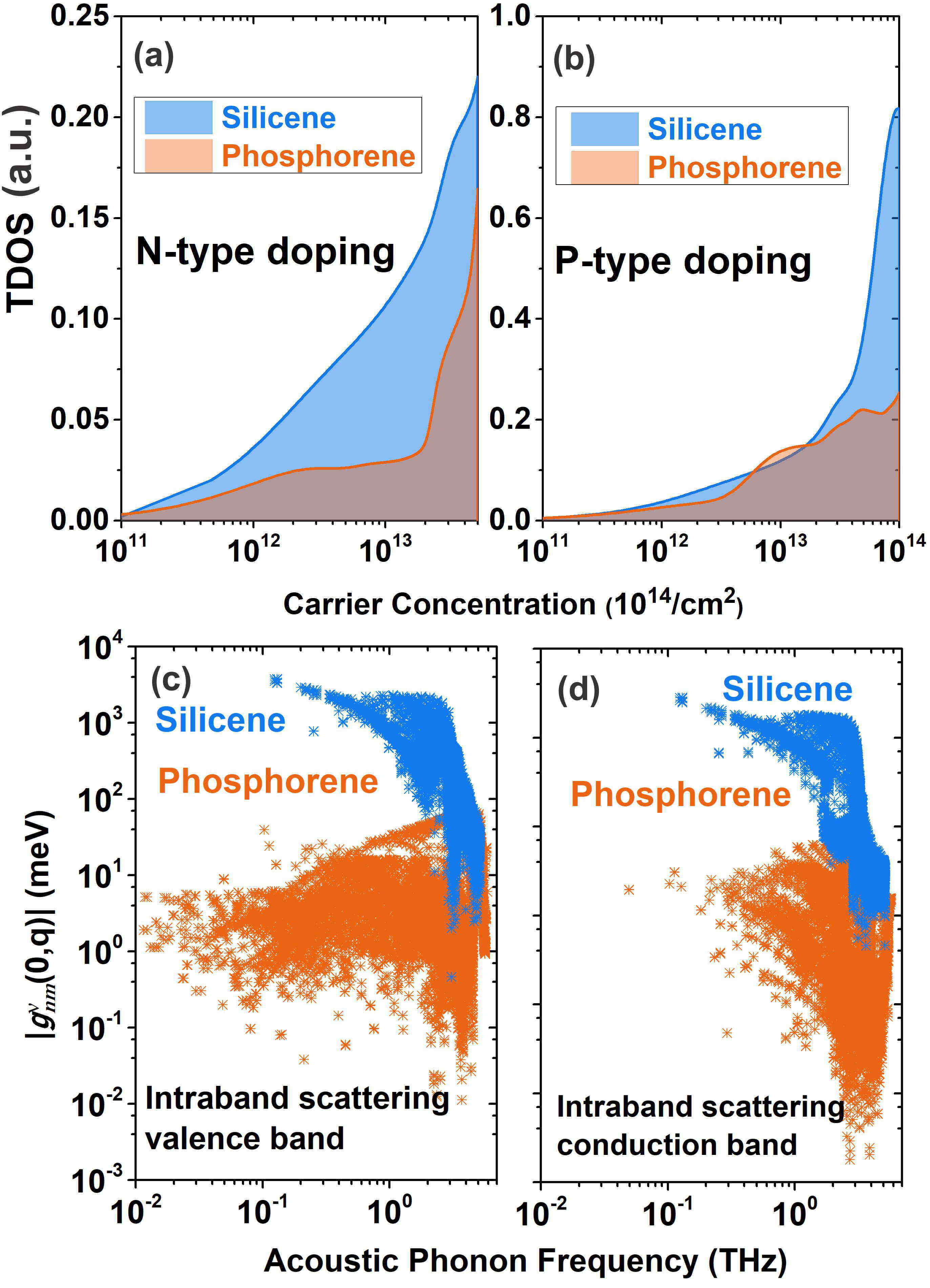}
\caption{The thermal density of states (TDOS) as defined in Eqn.~\ref{eqn:4} in (a) n-type and (b) p-type silicene and phosphorene. The intraband electron-phonon scattering matrix elements for acoustic phonons with the initial electron state at $\bf{k}=\bf{0}$ as a function of phonon frequency in (c) p-type and (d) n-type silicene and phosphorene. } 
\label{fig:4}
\end{figure}

From Figs.~\ref{fig:2} and \ref{fig:3}, it is clear that the phonon scattering rates due to phonon-electron scattering are significantly higher in silicene than those in phosphorene given similar carrier concentrations. To understand this difference, we analyze the strength of phonon-electron scattering by considering the two major factors that determine the phonon-electron scattering rates as reflected in Eq.~\ref{eqn:1}: the number of potential phonon-electron scattering channels imposed by the energy-momentum conservation conditions and the magnitude of the electron-phonon scattering matrix elements. Given the small energy scale of phonons compared to that of electrons, phonon-electron scatterings are approximately ``on-shell" processes, where the initial and the final electron states have similar energy. This observation implies that the number of potential phonon-electron scattering channels is directly related to the electron density of states within the Fermi window, where scatterings can happen between partially occupied energy levels. A quantitative measure is the so-called ``thermal density of states'' (TDOS)\cite{grimvall1981electron}, which is defined as a function of the Fermi level:
\begin{equation} \label{eqn:4}
    \mathrm{TDOS}(E_\mathrm{F})=\int_{-\infty}^{+\infty} \mathrm{DOS}(E)f'(E,E_\mathrm{F}) dE,
\end{equation}
where $\mathrm{DOS}(E)$ is the electron density of states, and $f'(E,E_\mathrm{F})$ is the energy derivative of the Fermi-Dirac distribution, which defines the Fermi window. In Fig.\ \ref{fig:4}(a) and (b), the TDOS for both n-type and p-type silicene and phosphorene are shown, which clearly signal that there is a larger available scattering phase space in silicene than phosphorene at the same carrier concentration due to silicene's semimetallic band structure. In Fig.~\ref{fig:4}(c) and (d), we further calculate and compare the scattering matrix elements for acoustic phonon modes in silicene and phosphorene given the same initial electron states at $\mathbf{k}=\mathbf{0}$ and at the band extrema. The electron-phonon matrix elements can be calculated as $g^{\nu}_{mn}(\mathbf{k},\mathbf{q})=\Braket{u_{m(\mathbf{k}+\mathbf{q})}|\Delta_{\mathbf{q} \nu} v^{\mathrm{SCF}} | u_{n\mathbf{k}}}$, where $u_{m(\mathbf{k}+\mathbf{q})}$ and $u_{n\mathbf{k}}$ are the periodic parts of the Bloch wavefunctions of the corresponding electron states and $\Delta_{\mathbf{q} \nu} v^{\mathrm{SCF}}$ is the perturbation of the electron self-consistent-field potential induced by a phonon mode with wavevector $\mathbf{q}$ and branch index $\nu$\cite{giustino2017electron}. Physically, the matrix elements measure the sensitivity of the electron potential energy in response to lattice disturbance and can be affected by crystal structure, chemical bonding environment and electron screening effect\cite{grimvall1981electron}. Here we see that the electron-phonon scattering matrix elements in silicene are significantly larger in magnitude than those in phosphorene, signaling that the electron states in silicene are much more sensitive to lattice perturbation and the structure of silicene might thus be less stable than that of phosphorene.

\begin{figure*}[h!]
\includegraphics[width=1\textwidth,clip]{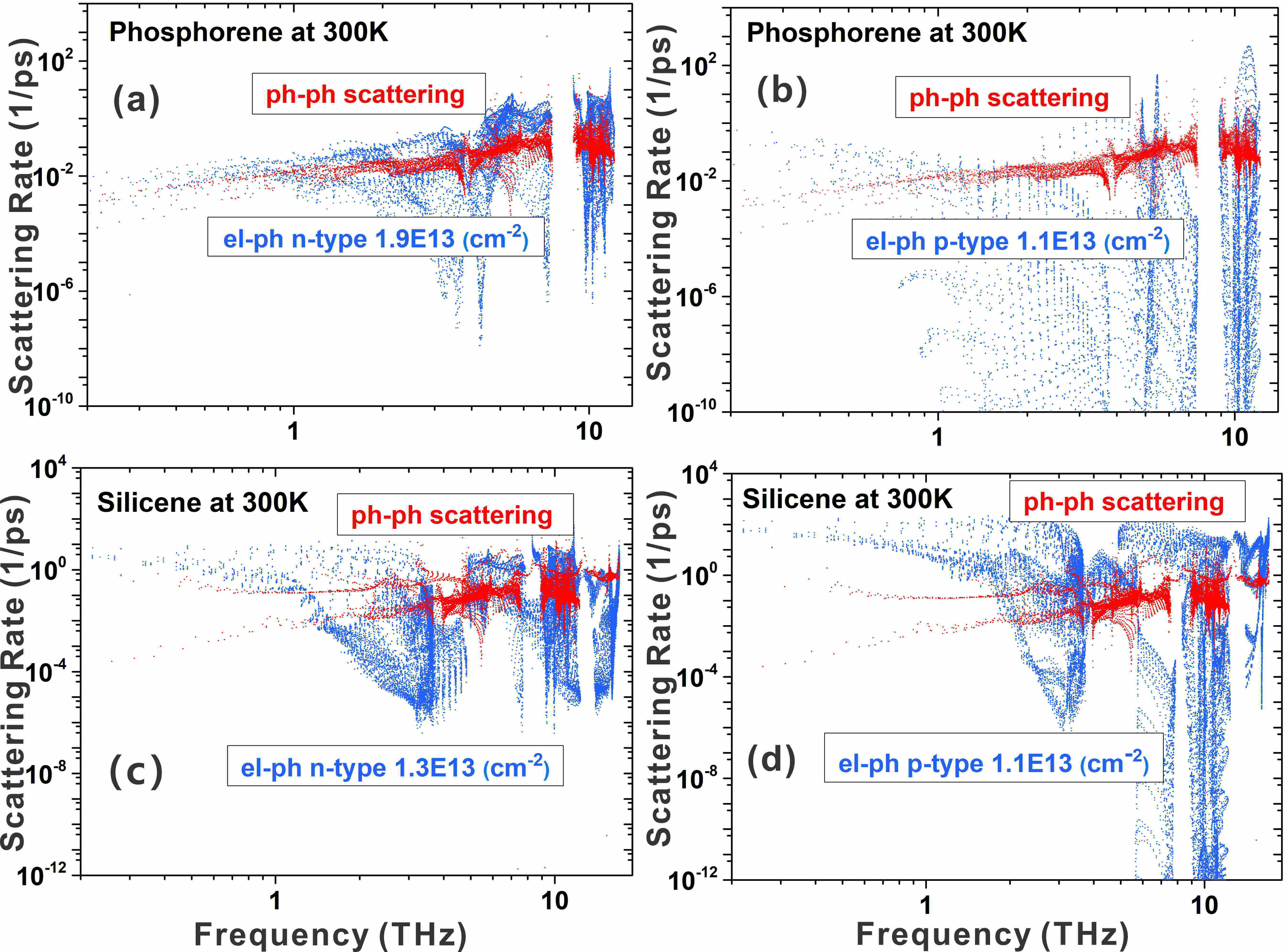}
\caption{Comparison of phonon-phonon and phonon-electron scattering rates as a function of phonon frequency in (a) n-type phosphorene, (b) p-type phosphorene, (c) n-type silicene and (d) p-type silicene.}
\label{fig:5}
\end{figure*}

Finally, to compare the contribution of phonon-electron scattering to the lattice thermal conductivity with that of intrinsic phonon-phonon scattering, we calculate the phonon-phonon scattering rates in silicene and phosphorene based on the anharmonic interatomic force constants (IFCs). The anharmonic IFCs are calculated using a frozen-phonon approach with supercell sizes $\rm 5\times 5\times 1$ and $\rm 4\times 4 \times 1$ for silicene and phosphorene, respectively. The q-mesh grid $60 \times 60 \times 1$ is adopted for both silicene and phosphorene. The results are shown in Fig.~\ref{fig:5}. When the carrier concentration is higher than $1 \times 10^{13}\ \mathrm{cm}^{-2}$, the scattering rates of the low-frequency acoustic phonons below a few THz are dominated by phonon-electron scattering, which is expected to have a major impact on the lattice thermal conductivity of these materials. To quantify this effect, we further combine the phonon-phonon and phonon-electron scattering rates using Matthiessen's rule and solve the phonon Boltzmann transport equation with the ShengBTE package\cite{shengbte01,shengbte02} to obtain the lattice thermal conductivity as a function of the carrier concentration, which is shown in Fig.~\ref{fig:1}(c). Due to the ambiguity of defining an effective thickness for 2D materials to calculate their thermal conductivity, here we report the relative reduction of the lattice thermal conductivity due to phonon-electron scattering normalized to the intrinsic values without net induced charge. The intrinsic thermal conductance (product of layer thickness and thermal conductivity) we obtain is $5.3 \times 10^{-9}\ \rm{W/K}$ for silicene, $5.8 \times 10^{-8}\ \rm{W/K}$ along the zigzag direction of phosphorene and $2.3 \times 10^{-8}\ \rm{W/K}$ along the armchair direction of phosphorene, which is in good agreement with literature values\cite{xie2014thermal,jain2015strongly}. Our results demonstrate that externally induced phonon-electron scattering can significantly affect thermal transport in silicene and phosphorene, particularly in silicene with high TDOS and strong electron-phonon coupling as reflected in the magnitude of the matrix elements. At a high carrier concentration of $10^{13} \ \rm{cm}^{-2}$, a \~40\% reduction of the lattice thermal conductivity in p-type silicene can be achieved, indicating the potential use of this mechanism to realize thermal switching devices. One potential concern is the increased electronic thermal conductivity as the induced carrier concentration increases. Due to the lack of available experimental data of electronic thermal conductivity in these materials, we estimate the electronic thermal conductivity using experimentally reported mobility\cite{tao2015silicene,liu2014phosphorene} and the Wiedemann-Franz law to be $\sim$ 10\% and $\sim$ 5\% of the intrinsic lattice thermal conductivity in silicene and phosphorene at a carrier concentration of $10^{13}\ \rm{cm}^{-2}$, respectively. It is worth noting that, in addition to inducing charge carriers, the external electric field applied to 2D materials can also cause changes to charge density distribution inside the material that can further reduce the thermal conductivity, as has been studied from first-principles in silicene\cite{qin2017external}. We envision that a combination of these effects will render silicene a promising candidate for thermal switching applications driven by an external electric field.

In summary, we analyze the impact of phonon-electron scattering on phonon transport in two representative 2D materials, silicene and phosphorene, using \textit{ab initio} calculations. We examine the mode-resolved phonon-electron scattering rates in silicene and phosphorene in detail and compare their behavior to 3D bulk materials. We explain the observed phonon-frequency dependence by developing a semi-analytical model using the deformation potential approximation, based on which we attribute the qualitatively different behavior to distinct phonon and electron dispersion relations in 2D materials as well as the reduced dimensionality. We provide further understanding about significantly different phonon-electron scattering rates in silicene and phosphorene by analyzing the magnitude of the scattering matrix elements and the available scattering phase space quantified by the thermal density of states. We find that the lattice thermal conductivity of silicene can be reduced by over 40\% via externally induced phonon-electron scattering, indicating the potential application in thermal switching devices. Our fundamental study also provides guidelines to search for 2D materials with even stronger phonon-electron scatterings for thermal switching applications.

\section{Author Contributions}
S. Y. and B. L. initiated this project. S. Y. carried out the \textit{ab initio} calculations. R. Y. developed the semi-analytical models. S. Y. and B. L. wrote the manuscript.

\begin{acknowledgement}
This work is based on research supported by the Army Research Office Young Investigator Program under the award number W911NF-19-1-0060. We acknowledge the use of the Center for Scientific Computing supported by the California NanoSystems Institute and the Materials Research Science and Engineering Center (MRSEC) at University of California, Santa Barbara through NSF awards DMR-1720256 and CNS-1725797. This work also used the Extreme Science and Engineering Discovery Environment (XSEDE) Stampede 2 at the Texas Advanced Computing Center (TACC) through allocation TG-DMR180044. XSEDE is supported by NSF grant number ACI-1548562.
\end{acknowledgement}

\bibliography{main}

\end{document}